\documentclass[]{aa}
\usepackage{epsfig}

\begin{document}

	\title{Dwarfs after Mergers? The case of NGC~520, NGC~772,
Arp~141, NGC~3226/7, NGC~3656 and Arp~299}

	\author{E. Delgado-Donate	
		\inst{1}
		\and
		C. Mu\~noz-Tu\~n\'on
		\inst{2}
		\and
		H. J. Deeg
		\inst{2} 
		\and
		J. Iglesias-P\'{a}ramo
		\inst{3}
		}

	\offprints{E. Delgado-Donate,\\
	 	\email{edelgado@ast.cam.ac.uk}
		}

	\institute{Institute of Astronomy, University of Cambridge,
	Madingley Road, Cambridge CB3 0HA, UK\\
		\email{edelgado@ast.cam.ac.uk}
		\and
		Instituto de Astrof\'{\i}sica de Canarias, E-38200 La Laguna, 
Tenerife, Spain\\
		\email{cmt@ll.iac.es, hdeeg@ll.iac.es}
		\and
		Laboratoire d'Astrophysique de Marseille, Traverse du
	Siphon - Les Trois Lucs, 13376 Marseille, France\\
              \email{jorge.iglesias@astrsp-mrs.fr}
		}

	\date{Received 21-06-02; accepted 13-02-03}

	\abstract{ We present results from a survey of dwarf galaxy
candidates in the vicinity of strongly interacting galaxies. The goal
of the survey was a test of the hypothesis that massive condensations
of stars and HI in tidal tails of large interacting galaxies may be a
significant source of independent, self-gravitating dwarf
galaxies. These so called {\it tidal dwarf galaxies} (TDG) can be
expected to resemble the blue luminous knots found in tidal tails, but
also might appear as redder, evolved systems if formed much before
they are being observed.  For the present study, a homogeneous
subsample of 6 fields was selected from the catalog of extended
objects in fields around 15 strongly interacting galaxies by Deeg et
al. (1998). Criteria for the subsample were: similar redshifts of the
central interacting galaxies, and photometric completeness of the
extended objects in V and R. The number density of these TDG
candidates was compared with expected background galaxy
densities. Within the statistical errors, background galaxies account
for most, if not all of the extended objects. There is no evidence for
a substantial locally formed dwarf galaxy population. Thus, we
conclude that field galaxy-galaxy interactions are likely to result in
the formation of only a few {\it long-lived} TDG.

		}

	\titlerunning{Dwarfs after Mergers?}

	\maketitle

\section{Introduction}

Galaxy interactions provide strong clues about the structure and
dynamics of galaxies, and a powerful mechanism for the origin of very
massive central star formation events. Besides, the development of
long tidal tails of stellar and gaseous material during the encounter,
which can be observed and computed in numerical simulations, has posed
the question of the fate of this tidally ejected material. Zwicky
(1956) was the first to propose that some of this material might
survive in the long term in the form of self-gravitating dwarf
galaxies. Evidence in favor of this scenario for the origin of dwarfs
comes from the identification of blue condensations in the tips of the
tails of galaxy mergers with typical H{\sc i} masses ranging from
$5\times 10^{8}$ to $6\times 10^{9}$ $M_{\odot}$.  Mirabel et
al. (1992) and Duc et al. (1998) have shown that objects similar to
dwarf irregulars or blue compacts are found in the tails. These small
galaxies of tidal origin are potential progenitors of detached
systems, namely, isolated dwarf galaxies (Sanders \& Mirabel 1996,
Braine et al. 2001).

Numerical simulations of interactions between disk galaxies show that
tidal tails are the natural result of such processes (Barnes \&
Hernquist 1996). Also, if the interaction is strong enough,
self-gravitating clumps of gas and stars are likely to form along the
tails (Schweizer 1978). Several examples of stellar condensations at
the tips of long tidal tails have been reported, like NGC~7252
(Hibbard et al. 1994), Arp~105 (Duc \& Mirabel 1994) and Arp~245 (Duc
et al. 2000). The possibility that these clumps become independent
dwarf companions has been proposed by Barnes \& Hernquist (1992) and
Elmegreen et al. (1993). In addition, some material -- both gaseous
and stellar -- is ejected from the disks of the parent galaxies,
forming lower mass complexes that are probably not
self-gravitating. It is expected that only the outermost ejected
material may eventually abandon the potential well of the parent
galaxy and form dwarf galaxies. The innermost material would slowly
fall back to the parent galaxy. Given that the further out the
material is, the longer is the time for return, dwarf objects in the
surroundings of the interacting pairs would be expected for a long
time, even well after the relaxation time. These self-gravitating
condensations of stars and H{\sc i}, the so-called Tidal Dwarf
Galaxies (henceforth TDG), would fade significantly in optical
wavelengths in the Gyr that follows their formation, as soon as the
gas is appreciably exhausted by the starburst (Weilbacher et
al. 2000). In addition, tidal stripping from the parent galaxy may be
in the long term a significant source of mass loss, thereby reducing
the TDG to a very low surface brightness dwarf (Mayer et
al. 2001). All these circumstances conspire to make genuine dwarf
galaxies hard to detect, unless H{\sc i} cartography as well as
spectroscopic surveys around mergers are performed.
 
In the present study we investigate if strong interactions between
galaxies can produce a significant enhancement on the number of dwarf
galaxies around the system. Lacking spectroscopic data, which are very
difficult to obtain due to the faintness and small size of the large
majority of TDG candidates, we choose to address this question
statistically, by comparing the photometric sample of extended objects
- or dwarf candidates- catalogued in the previous work of Deeg et
al. (1998, hereafter Paper~I) with background galaxy number counts. By
doing so, we expect to include both young and evolved TDG, arriving at
an estimate of the abundance of TDG formed by field interacting
galaxies.  From the fields around 15 interacting systems catalogued in
Paper~I, we have selected those fields that were observed in two
filters ($V$ and $R$), had a statistically significant number of
detections of extended objects (EO), and in which the central galaxies
have similar redshifts (five of the sample galaxies have v$_{\rm r}$ between
2266 and 3132 km s$^{-1}$, NGC~3226/7 is at 1151 km s$^{-1}$). Thus, a
subsample of fields around six galaxies -- namely NGC~520, NGC~772,
NGC~3226/7, NGC~3656, Arp~141 and Arp~299 - is analyzed in this
paper. Among these, Arp~299 was observed in one filter only, but the
high number of detections of EO in this field lead to its inclusion in
this study.

In the following, we describe the observations and the detection
procedure in Sect.~2.  In Sect.~3, we outline the properties of known
TDG within the framework of a dynamic evolution scenario and describe
the statistical analysis that was performed in order to quantify the
possible number of TDG. Conclusions are given in section Sect.~4.

\section{Observations and Detection of Candidates \label{obser}}

Images of the fields were taken in December 1992 with the 2.5m Isaac
Newton Telescope at the Observatorio del Roque de los
Muchachos, La Palma, with the
EEV5 CCD-camera in prime focus, giving an
effective field of about 10.5$~\arcmin \times 10.5~\arcmin$ and 0.55
$\arcsec$~pix$^{-1}$ resolution. The observations took place in
bright-grey time which affected the limiting
magnitude.  Atmospheric transparency conditions were
very good and seeing ranged from 0.9 to 1.5 $\arcsec$.  Images were
taken with the $V$ and $R$ Cousins filters, with exposure times as given in
Table~1.  In order to determine the stellar PSF, frames with short 20 s exposures were also taken, in which known bright
field stars are not saturated. Landolt (1992) standards were
observed for flux calibration.

\begin{table*}
\begin{center}
\begin{tabular}{lcclccccc} \hline
Galaxy & $\alpha$ & $\delta$ & Type & v$_{\rm r}$ & 
Gal. Size & t$_{\rm R}$ & t$_{\rm V}$ & Comm.\\  
 & & & & km s$^{-1}$ & $\arcmin^2$ & s & s & \\  
\hline
NGC~520	& 01:21:52 & 03:34:56 & S pec  & 2266 & 1.9$\times$0.7 & 300
& 600 & Arp~157 \\
NGC~772 & 01:56:26 & 18:45:05 & SA(s)b & 2472 & 7.2$\times$4.3 & 300
& 500 & Arp~78 \\
Arp~141 & 07:08:12 &73:33:56 & E       & 2735 & -- & 500 &
600 & -- \\
NGC~3226/7 & 10:20:47 & 20:07:00 & E2:pec & 1151 & 3.2$\times$2.8 & 600
& 600 & Arp~94\\
NGC~3656  & 11:20:50 & 54:07:06 & (R')I0:pec & 2860 & 
1.6$\times$1.6 & 600 & 600 & Arp~155\\
Arp~299 & 11:25:42 & 58:50:14 & SBm pec & 3132 & 1.2$\times$1.0 & 200
& -- & NGC~3690\\
\hline
\end{tabular}
\caption{Physical and observational parameters of the sample.}
\end{center}
\end{table*}

Table~1 lists relevant parameters of the target list and of the observations.

Raw images were zero-subtracted and flatfielded in the usual
manner. The bias frame used was taken from averaging all bias frames in the
observations. Also, {\em global} flatfields were obtained by averaging
the individual sky flatfields taken at the entire observing run. This was
possible, since the stability of the flatfields between nights was better
than 1\%. 
An absolute flux calibration was performed based on the observations
of Landolt standards at different airmasses.  The standards were
measured using fixed-aperture photometry.  For more details about
observations, calibration procedure and classification methodology, see Paper~I. A procedure was developed to identify objects over the background of the
CCD frames, discriminating small galaxy-like objects and star-like
objects (see the flow diagram in Paper~I).
Identified objects were then parameterized by the fitting of elliptical
Gaussians, as is described later in the text. The total error in the
photometric calibration is about 0.15~mag.  

\begin{table}
\begin{center}
\begin{tabular}{lccc}
\hline
Field &  \# EOs & \# SOs &\# amb. \\  \hline
NGC~520  &   &   &   \\  
all & 85 & 150 & 44\\
$R \leq 20.0$ & 45 & 70 & 17\\	 \hline
NGC~772  &   &   &   \\  
all & 119 & 209 & 68 \\
$R \leq 20.5$ & 84 & 98 & 27\\	 \hline
Arp~141  &   &   &   \\  
all & 183 & 623 & 109 \\
$R \leq 20.5$ & 76 & 301 & 27\\	 \hline
NGC~3226/7  &   &   &   \\  
all & 156 & 260 & 71 \\
$R \leq 20.5$ & 63 & 91 & 10\\	 \hline
NGC~3656  &   &   &   \\  
all & 257 & 510 & 152 \\
$R \leq 20.5$ & 76 & 72 & 16\\	 \hline
Arp~299  &   &   &   \\  
all & 116 & 181 & 74 \\
$R \leq 20.5$ & 81 & 76 & 25\\	 \hline
\end{tabular} 
\caption{Number of Extended (EO), Stellar (SO) and
ambiguous objects (amb) identified in the field of each of the
galaxies. Shown are counts for all identified objects, as well as
counts for objects within the limiting magnitude for completeness of the catalogue.}
\end{center}
\end{table}

Objects with a size close to the PSF were considered stellar (SO)
whereas those with an extent of more than 1.3 times the PSF in either
$x$ or $y$ direction were classified as extended (EO).  Objects whose
FWHM ranges between 1 and 1.3 times the PSF were classified as
ambiguous.  All saturated objects, also those having widths larger
than the PSF, can safely be classified as stellar, too.  Table~2 gives
the number of EOs, SOs and ambiguous identifications found around the
galaxies of the sample.  The selection method used certainly does not
exclude that the lists of galaxies still include some semi-resolved
double or multiple stars. Also the lists of stars may contain a
substantial number of galaxies with a near--stellar PSF.  In
galaxy--star separations done in a way comparable to ours, Metcalfe et
al. (1991) found that later spectroscopy revealed 29\% of the
star--like objects to be compact galaxies, and 5\% of the galaxy
identifications turned out to be stars.

It might be argued that some of the detected objects are globular
clusters formed during the interaction. In fact, Zepf \& Ashman (1993)
and Ashman \& Zepf (1992) reported two-aged populations of globular
clusters around merger remnant candidates. The colors of our objects
would be compatible with those of typical globular clusters (Harris
1996), thus opening the possibility that they are globular clusters
that had originated during the interaction. However, globular clusters
should be detected as point-like sources at the distances of the
interacting galaxies of our sample. In addition, only the faintest EOs
in our sample would have luminosities comparable to those of globular
clusters.

\begin{figure*}
\begin{center}
\mbox{\epsfig{file=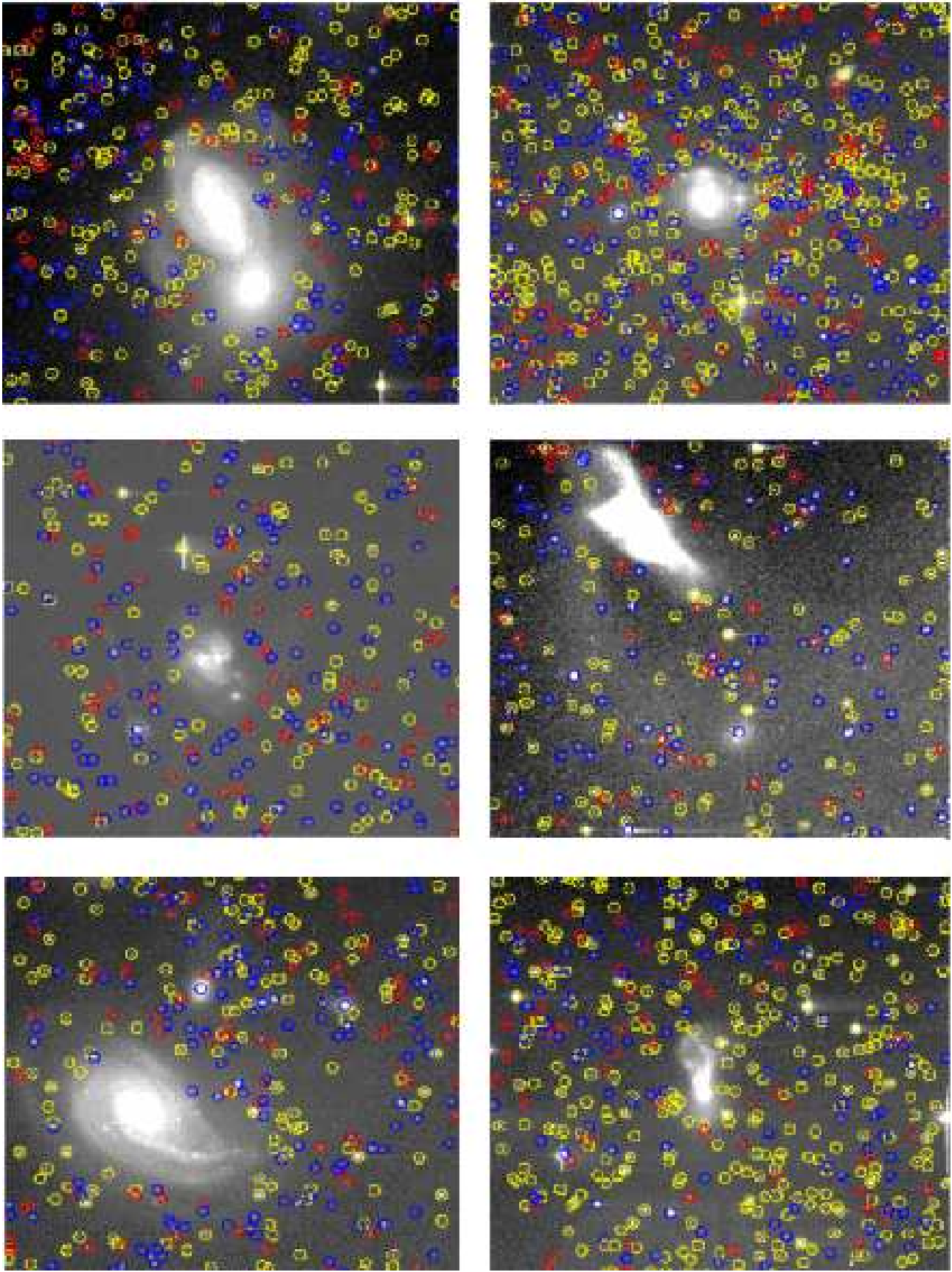,height=23cm}}
\caption{R band CCD image of NGC~3226 (top left), NGC~3656 (top
right), Arp~299 (center left), NGC~520 (center right), NGC~772
(bottom left) \& Arp~141 (bottom right). Light blue, yellow and red open
circles represent detected EO, SO and ambiguous objects
respectively. South is at the top and East towards the left.\label{mosaic}}
\end{center}
\end{figure*}

Figure~1 shows calibrated $R$ band images of the sample fields, in
which the EO, SO and ambiguous objects that have been detected are
marked. Below, we give some comments on the galaxy fields:

\begin{itemize}
\item NGC~520: This is a gas rich system that is widely studied in the
literature. An interesting multifrequency analysis of this interacting
system can be found in Hibbard \& van Gorkom (1996). Two dwarf objects
with radial velocities similar to that of the parent galaxy,
detected in both the optical and in H{\sc i}, are present in the
neighborhood of the merger, and one of them (UGC957) appears  {\it
permeated} by the
northern tidal tail of the system. This latter object might be the result of
the interaction. 
\item NGC~772: Clear disruption signatures can be found in
the spiral arms of this galaxy, which are likely due to interactions with
some of the satellites detected around it (see Zaritsky et al. 1997).
\item Arp~141 (UGC~3730): This is quite an irregular system with an
optical extension towards the South. Burbidge \& Burbidge (1959)
attributed the peculiar nature of this interacting pair to the
disruption of a spiral galaxy by an elliptical one in a close
encounter.
\item NGC~3226/7: This peculiar system is composed by NGC~3226 and
NGC~3227. The galaxies are embedded in an extended diffuse
envelope. H{\sc i} was detected along the two plumes emerging from
NGC~3227 (see Mundell et al. 1995).
\item NGC~3656: This shell galaxy was studied in detail by Balcells
(1997). Two low surface brightness tidal tails were reported. At the tip
of the western tidal tail, two dwarf galaxies were detected in broad
band images. However, it is not certain that these objects are associated with NGC~3656.
\item Arp~299: This system consists of two interacting spiral
galaxies, IC~649 and NGC~3690. Multi-wavelength HST observations of
NGC~3690 highlight two
massive star--forming condensations $\approx$~3~kpc from the merger
nucleus that might become detached dwarf galaxies (Alonso--Herrero et
al. 2000). An H{\sc i}--rich tidal
tail towards the North is apparent in larger optical frames. Surprisingly,
no dwarf galaxies were detected along this tail (Hibbard \& Yun,
1999).
\end{itemize}

\section{Results
	\label{results}
	}

\subsection{TDG properties and the collision process}

\begin{figure*}  
\begin{center}
\mbox{\epsfig{file=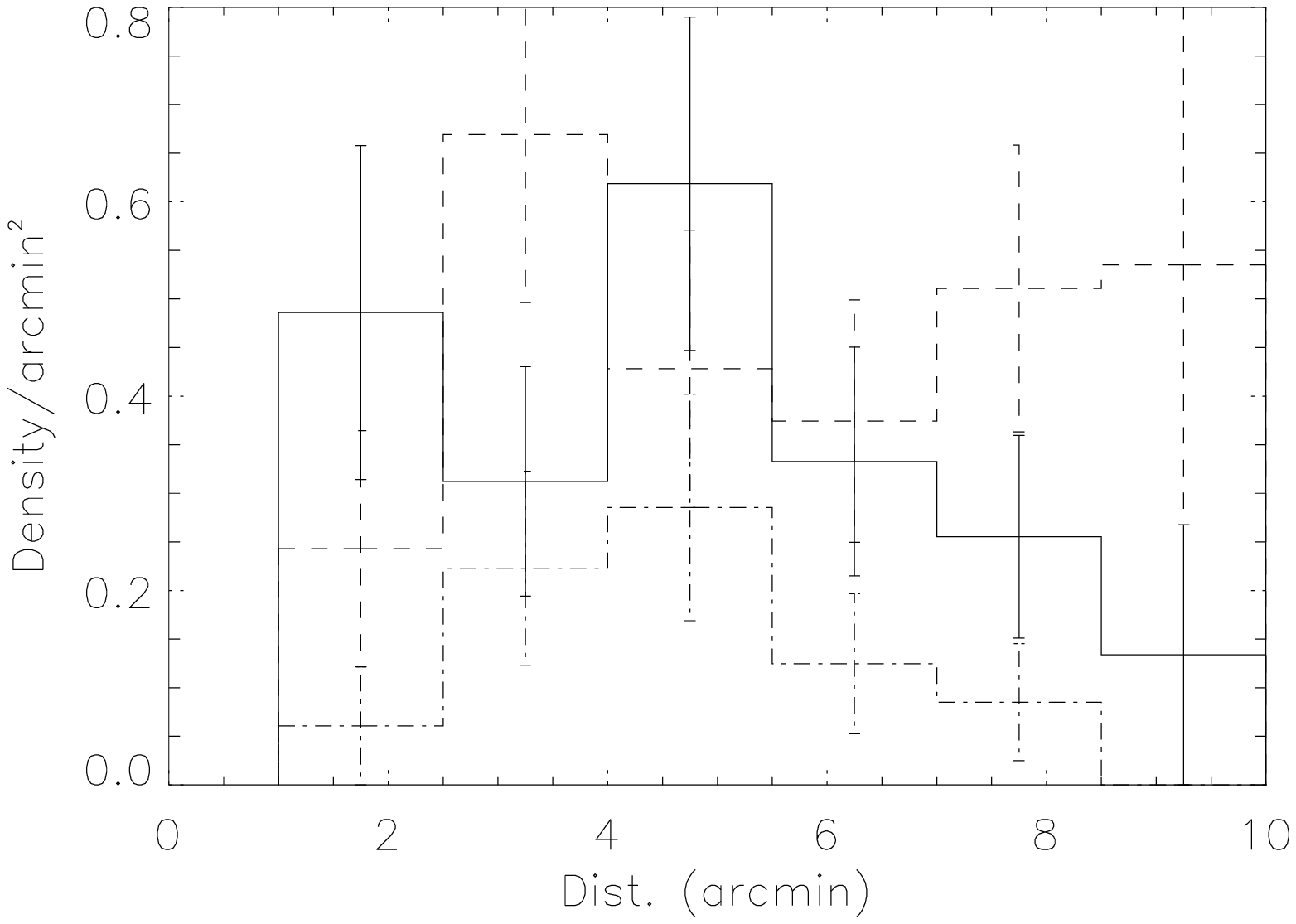,height=6cm}\psfig{file=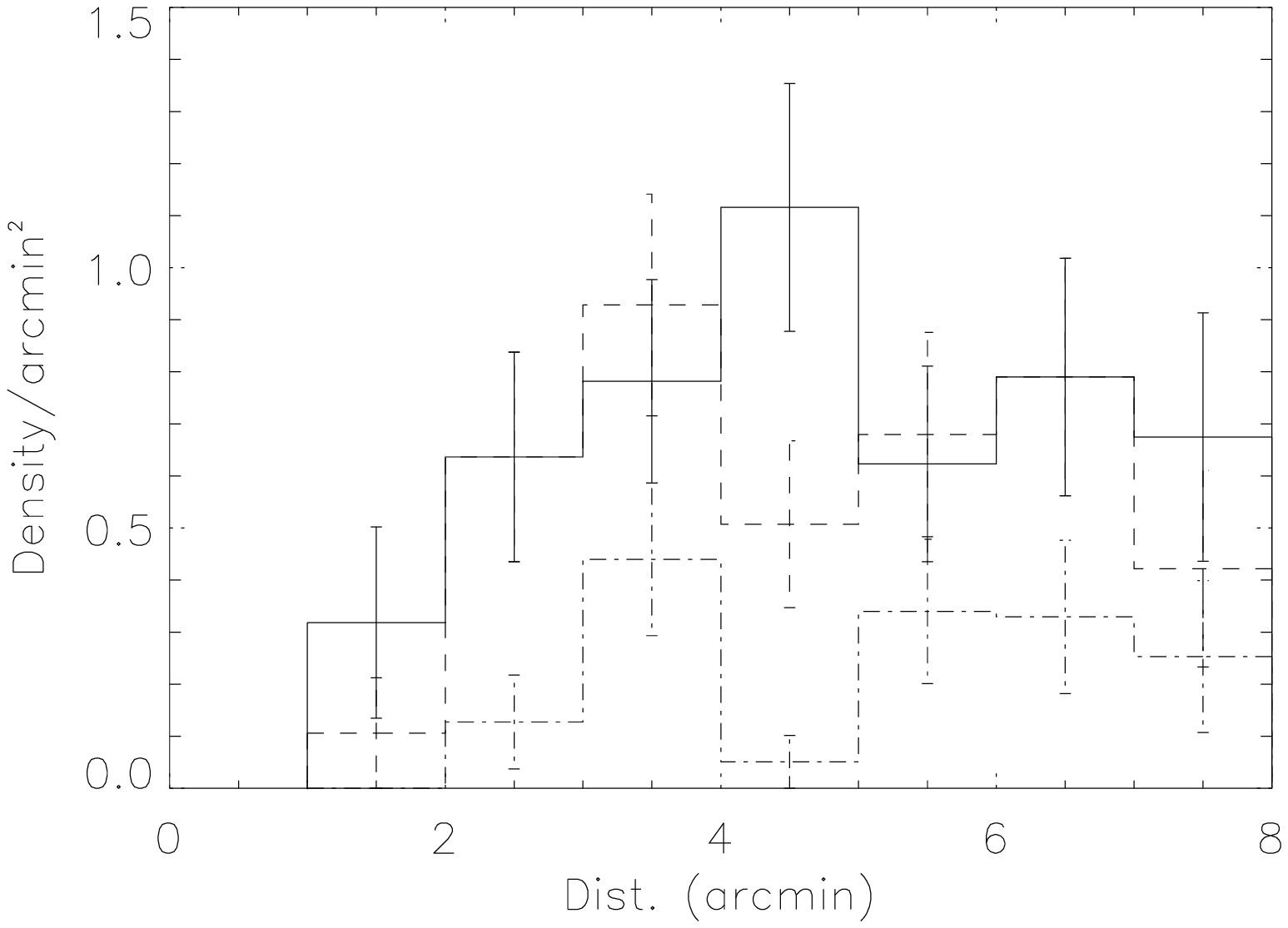,height=6cm}}
\mbox{\epsfig{file=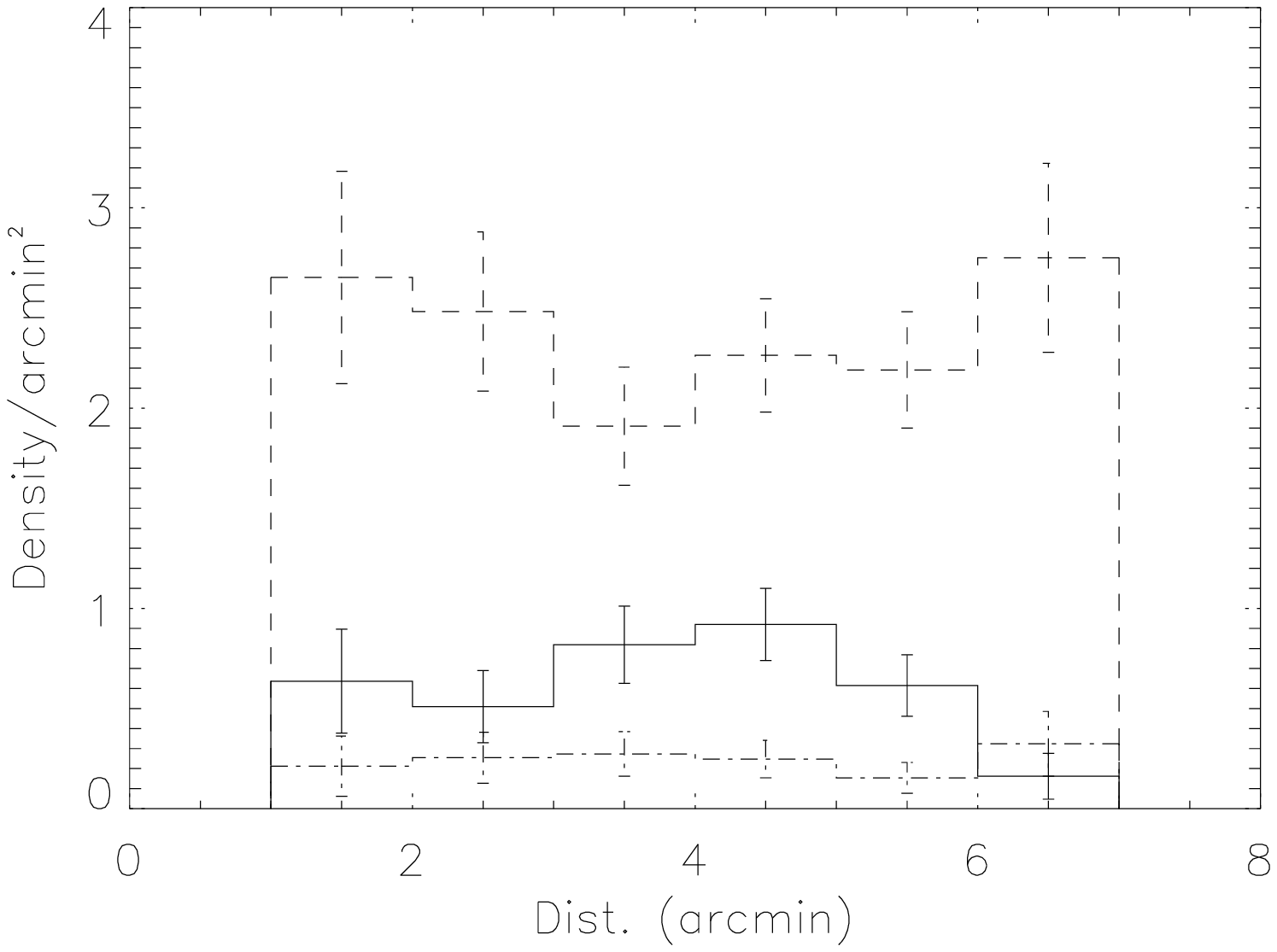,height=6cm}\psfig{file=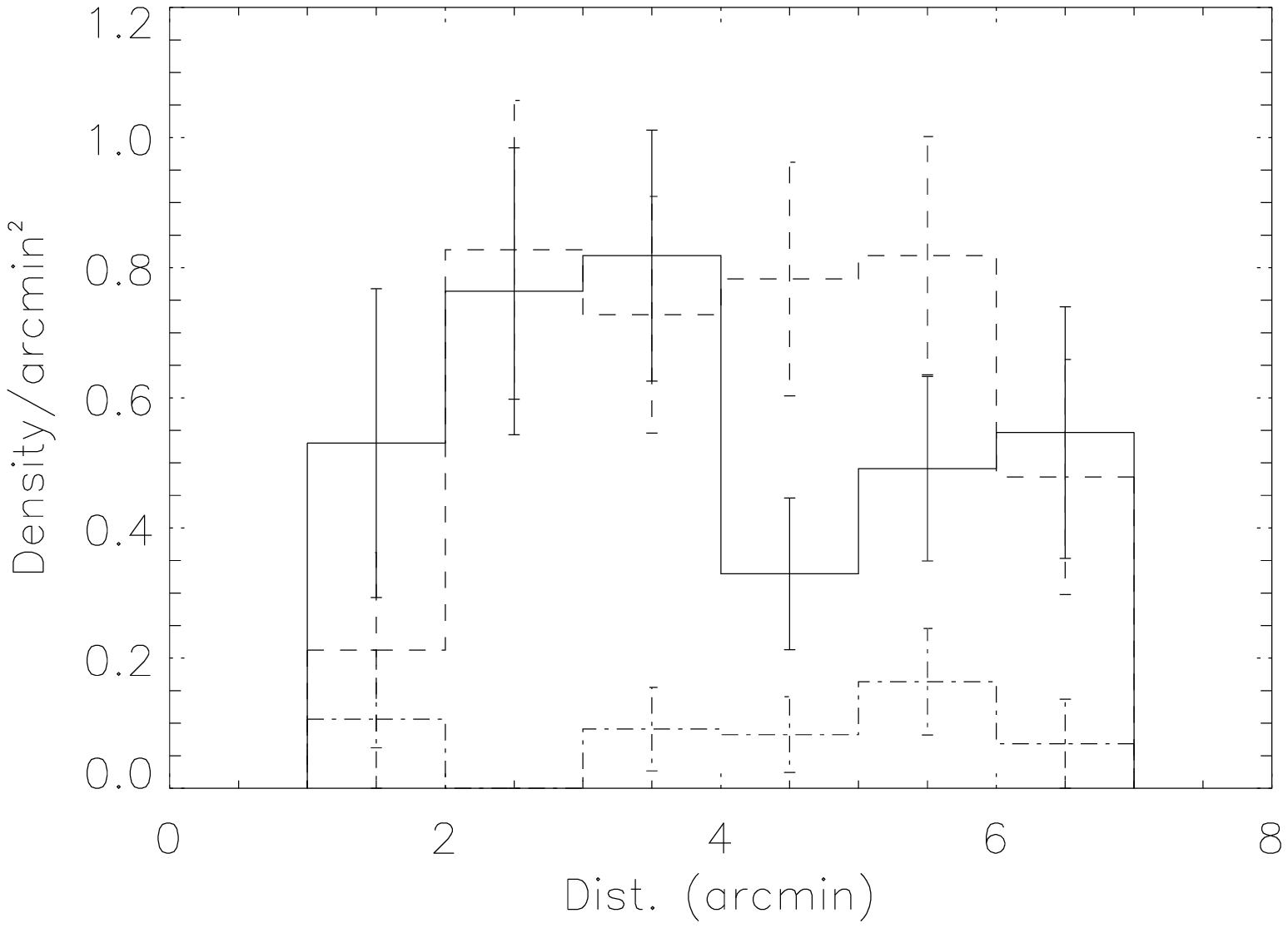,height=6cm}}
\mbox{\epsfig{file=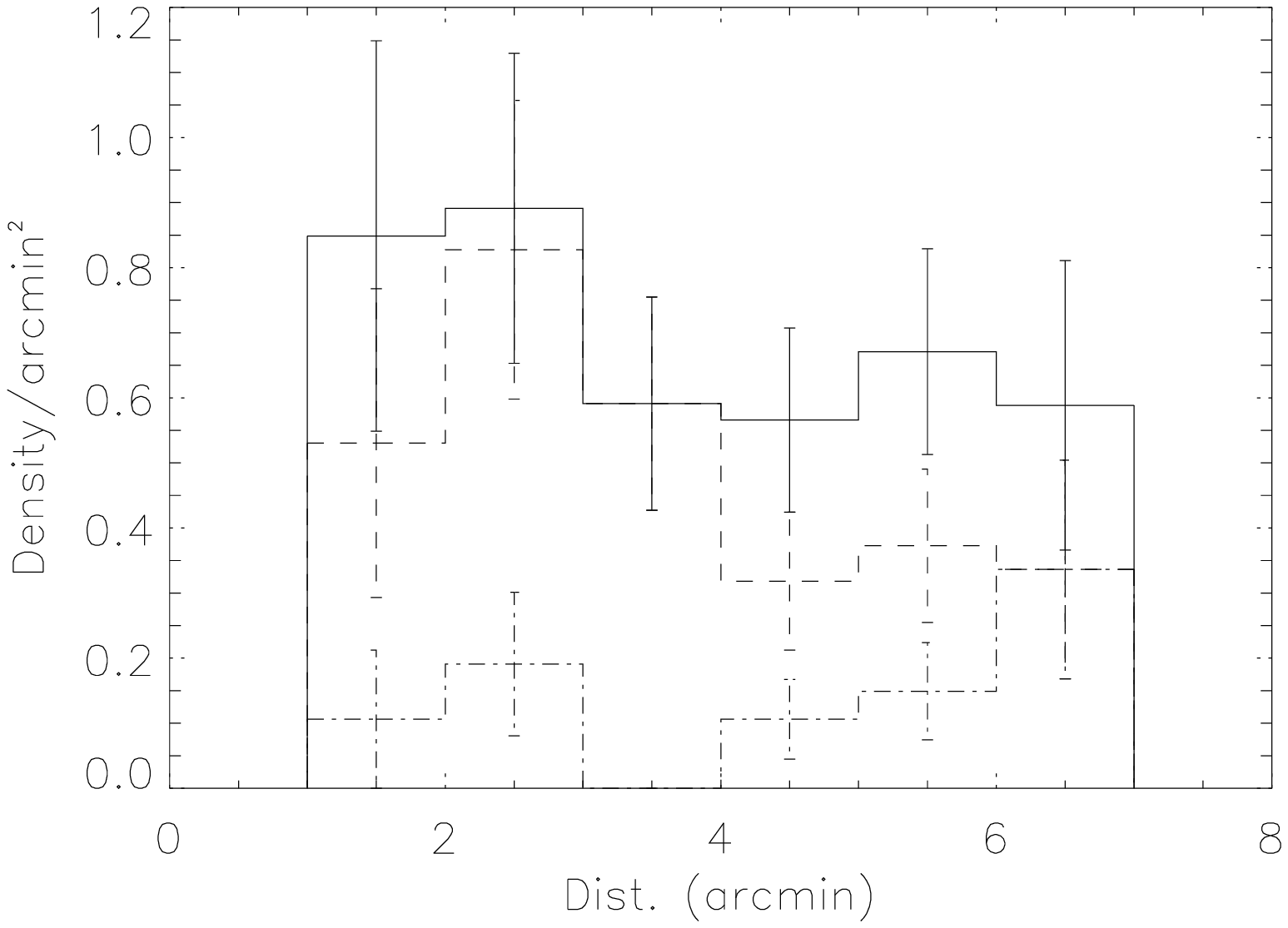,height=6cm}\psfig{file=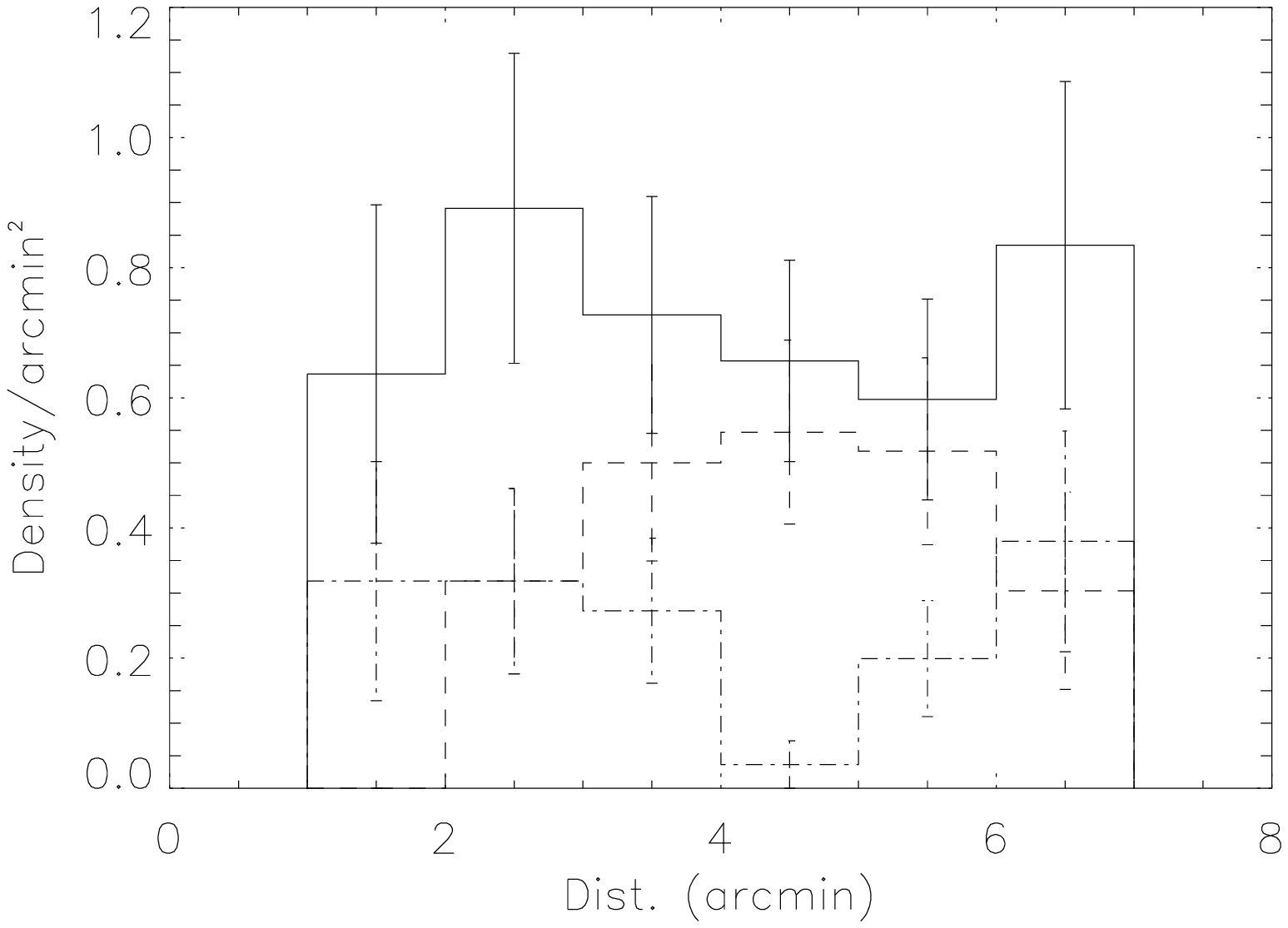,height=6cm}}
\caption{ Density Distribution of objects versus  distance for each
field. The solid line represents EOs, the dashed
line SOs and the dot-dashed line ambiguous objects. From left to
right: [top] NGC~520 \& NGC~772; [center] Arp~141 \& NGC~3226/7;
[bottom] NGC~3656 \& Arp~299. The plots include all galaxies detected 
within the completeness limit (see Table~2). }
\end{center}
\end{figure*}

TDGs, {\it objects which are self-gravitating entities, formed out of
the debris of a galactic gravitational interaction} (Duc et al. 2000)
are characterized by a luminosity comparable to that of typical dwarf
galaxies, an extended morphology, high metallicity, and blue colors
($B-V\approx 0.5$ and $V-R\approx 0.4$) as a result of an active
starburst. These properties, however, have been outlined for objects
found at the tip of the optical tails, some of them kinematically
decoupled from the surrounding material but still {\it visually}
embedded in the collision debris. On the other hand, more numerous (
$\approx 4$ TDG per collision) TDG candidates have been found, both in
broad band optical images (Weilbacher et al. 2000) and in H$\alpha$
emission (Iglesias-P\'aramo \& V{\'{\i}}lchez 2001), distributed all
along the tidal features. However, it remains unclear whether these
structures might become truly detached dwarfs in the future. How many
TDG does an average merger form? Computer simulations of encounters
between equal-mass disk galaxies (Barnes, private communication) show
the formation of $\approx 100$ {\it bound structures} along the tidal
tails, with a distribution of distances from the merger that is very
broad and evolves with time as the tail expands. Only a few bound
systems have masses above $5\times10^{8}$ $M_{\odot}$ and they are
located randomly throughout the tail, at an approximate mean distance
of 100~kpc. Among these most massive self-gravitating structures,
those with larger radii are the best candidates for becoming
independent dwarf galaxies in the long term (Barnes \& Hernquist 1992,
Elmegreen et al. 1993). Most of the other bound structures would fall
back to the central galaxy within a Gyr.  One should however consider
that in the case of interactions between unequal mass galaxies, the
length of the tidal tail should be much smaller than in the case
reported above. Correspondingly, dwarf candidates would be located
much closer to the nuclei of the interacting pair.

One might imagine a dynamic scenario in which, depending on the nature
(morphology, mass) of the galaxies involved in the collision and the
stage at which the interaction is being observed, TDG do not
necessarily have the properties observed in their young
counterparts. TDGs could have formed, developed a massive starburst,
and faded away, as well as decoupled from the main stream of merger
debris -- Weilbacher et al. 2002 reports on 7 {\it knots} along
the tails of AM~1353~272 which could be moving away from the parent
tidal stream with speeds as large as 100 km~s$^{-1}$. In such a
picture, it would be possible to find TDGs without blue {\it excess}
located well away from the tidal features, in numbers that would be
hard to predict. Weilbacher et al. (2000) have computed the future
luminosity evolution of the tidal knots detected in a sample of 10
interacting galaxies, assuming that a single starburst episode
occurs. The colors become significantly bluer (at first mainly in
$V-R$) once the burst commences, and redden again in the next Gyr, as
the stars formed in the burst age. Therefore, TDGs around interacting
galaxies may comprise a large population, perhaps {\it hidden} among
background faint galaxies in the neighborhood of the interacting
pair. A series of tests to assess this scenario was undertaken and is
described in the next section.

\subsection{Analysis}

\begin{figure}  
\begin{center}
\mbox{\psfig{file=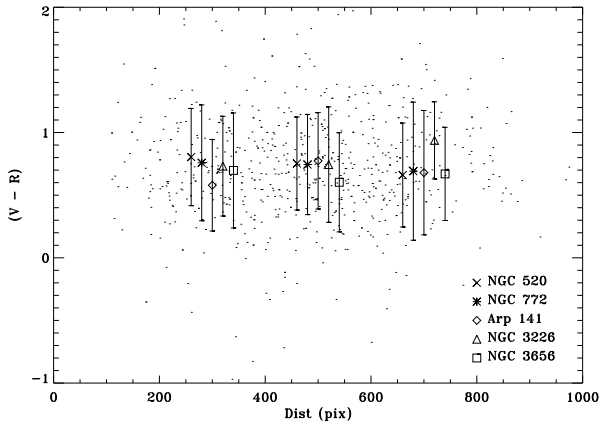,height=6cm}}
\caption{Dependence of the $(V - R)$ color index of the detected
objects as a function of spatial distance in pixels (1 pixel = 0.55
arcsec) to the parent galaxies' center. The symbols and error bars show the
averaged colors over different
radial distances. Arp~299 is not included since no colors are available.}
\end{center}
\end{figure}

We have computed the number density distribution $\mu$ of
EOs and SOs in each field (Fig.~2). This quantity has been used
to look for a possible dependence of $\mu_{\rm EO}$ and $\mu_{\rm SO}$ with
distance from the parent galaxies, whose center was taken as reference
point (RP). Each CCD frame was 
divided in concentric rings around the RP. Only rings close to the
center are completely present on the frames. To include rings
further out, an area-normalization algorithm was employed that takes
the irregular shapes of incomplete rings into account. Care was taken,
through a revision of the flatfield frames, that vignetting is negligible, as
it would cause lower detection limits far from the optical axis.

\begin{figure}  
\begin{center}
\mbox{\psfig{file=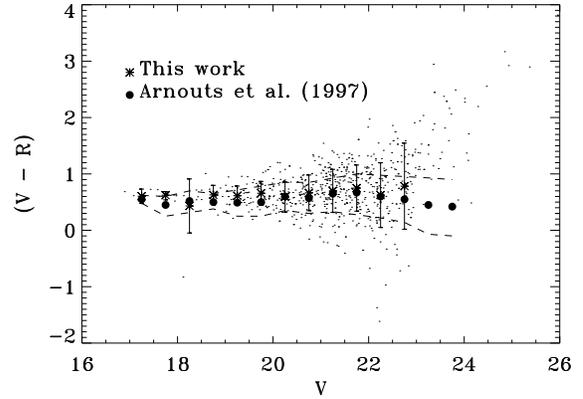,height=6cm}}
\caption{$(V - R)$ color index as a function of $V$
magnitude. Crosses with
error bars indicate the averaged V-R values per half V magnitude, with the corresponding standard deviation. Over-plotted are the mean values (filled circles) and
$1\sigma$ standard deviation of the colors (dashed lines) for the
photometric sample of background galaxies reported by Arnouts et al. (1997).}
\end{center}
\end{figure}

As can readily be seen, there is no clear evidence for a
characteristic distance where EOs are preferentially appearing.
Within the error bars, the density distributions are essentially
flat. Note that the last bin includes the outermost EO detected near
the edges of the CCD frames and their small numbers produce
unreliable statistical averages. This result does not depend on
color. In addition, as is expected for a random spatial distribution,
EO and SO distributions do not show any statistically significant
difference.

Figure~3 also shows the distribution (small points) of the radial
distance of the detected EO as a function of the $(V - R)$ color
index. The thick symbols stand for the average values over a given
radial distance interval.  Error bars correspond to the standard
deviation. Overall, the averaged values are -- within the error bars
-- independent of the distances to the central objects. Therefore, we
conclude that the color distribution of EOs is fairly homogeneous,
with no privileged position or distance in any of the studied fields.

Figure~4 shows the $(V - R)$ color index as a function of the $V$
magnitude for the EOs detected in all the frames. Crosses with error
bars correspond to the averaged values per half unit magnitude and one
standard deviation respectively. For comparison, we have over-plotted
the mean values (filled circles) and $1\sigma$ standard deviation of
the colors (dashed lines) for the photometric sample of background
galaxies reported by Arnouts et al. (1997).  As can be seen, both sets
of averaged values agree very well with each other. There is therefore
no indication for a foreground population with a distinct color.

Finally, the number of EOs in each field, in intervals of
0.5~$R$~magnitudes as well as over the entire magnitude range, was
computed and compared with the number of expected background galaxies
(Table~3). The purpose of the binning is twofold: first, it is
required to compute the errors in the number counts consistently and,
second, it allows us to search for possible excesses over background
counts within specific magnitude ranges. The abundances of expected
background galaxies were calculated from galaxy counts by Metcalfe et
al (1991). Possibly, galaxy counts from observations of nearby blank
fields might have added support to these results.  However, due to the
similarity between our star-separation procedure and the one used by
Metcalfe et al. (1991 -- both are based on a source size criterion) we
are confident that this would not have changed the results
significantly.

\begin{table*}
\begin{center}
\begin{small}
\begin{tabular}{lcccccccc}
\hline
 & N & $\sigma$ & N$_{\rm e}$ & $\sigma_{e}$ & dN $\pm$ $\sigma_{\rm d}$ &
 dN~deg$^{-2}$~mag$^{-1}$ &  $\sigma_{\rm d}$~deg$^{-2}$~mag$^{-1}$ & N$_{\rm max}$\\ \hline
\multicolumn{9}{c}{NGC~520}\\
$R \leq 18.0$ & 4 & 3 & 7 & 4 & {\bf ~-3 $\pm$ 7} & -197 & 432 & 4 \\
$18.0 < R \leq 18.5$ & 5 & 2 & 4 & 1 & {\bf ~~1 $\pm$ 3} & 63 & 194 & 4 \\
$18.5 < R \leq 19.0$ & 11 & 4 & 5 & 1 & {\bf ~~6 $\pm$ 5} & 386 & 286 & 11 \\
$19.0 < R \leq 19.5$ & 10 & 2 & 11 & 3 & {\bf ~-1 $\pm$ 5} & -34 & 286 & 4\\
$19.5 < R \leq 20.0$ & 15 & 3 & 23 & 5 & {\bf ~-8 $\pm$ 8} & -510 & 519& 0\\ 
$ All ~~R$ & ~ & ~ & ~ & ~ & {\bf ~-5 $\pm$ 13} & ~ & ~ & {\bf +8}\\
\hline
\multicolumn{9}{c}{NGC~772}\\
$R \leq 18.0$ & 12 & 7 & 6 & 4 & {\bf ~~~6 $\pm$ 11} & 387 & 744 & 17 \\
$18.0 < R \leq 18.5$ & 5 & 1 & 3 & 1 & {\bf ~~2 $\pm$ 2} & 104 & 173 & 4 \\
$18.5 < R \leq 19.0$ & 17 & 5 & 4 & 1 & {\bf ~13 $\pm$ 6} & 892 & 438 & 19 \\
$19.0 < R \leq 19.5$ & 13 & 2 & 9 & 2 & {\bf ~~4 $\pm$ 4} & 256 & 324 & 8 \\
$19.5 < R \leq 20.0$ & 18 & 3 & 21 & 5 & {\bf ~-3 $\pm$ 8} & -178 & 574 & 5 \\
$20.0 < R \leq 20.5$ & 19 & 2 & 32 & 6 & {\bf -13 $\pm$ 8} & -909 & 599 & 0 \\
$ All ~~R$ & ~ & ~ & ~ & ~ & {\bf ~+9 $\pm$ 17} & ~ & ~ & {\bf +26}\\
\hline
\multicolumn{9}{c}{Arp~141}\\
$R \leq 18.0$ & 6 & 4 & 7 & 4 & {\bf ~-1 $\pm$ 8} & -65 & 503 & 7 \\
$18.0 < R \leq 18.5$ & 2 & 1 & 4 & 1 & {\bf ~-2 $\pm$ 2} & -115 & 110 & 0 \\
$18.5 < R \leq 19.0$ & 9 & 3 & 4 & 1 & {\bf ~~5 $\pm$ 4} & 282 & 254 & 9 \\
$19.0 < R \leq 19.5$ & 15 & 4 & 10 & 2 & {\bf ~~5 $\pm$ 6} & 297 & 397 & 11 \\
$19.5 < R \leq 20.0$ & 16 & 3 & 23 & 5 & {\bf ~-7 $\pm$ 8} & -420 & 541 & 1 \\
$20.0 < R \leq 20.5$ & 28 & 5 & 35 & 7 & {\bf ~~-7 $\pm$ 12} & -470 &
 739 & 5 \\ 
$ All ~~R$ & ~ & ~ & ~ & ~ & {\bf ~-7 $\pm$ 18} & ~ & ~ & {\bf +11}\\
\hline
\multicolumn{9}{c}{NGC~3226/7}\\
$R \leq 18.0$ & 5 & 3 & 7 & 4 & {\bf ~-2 $\pm$ 7} & -117 & 482 & 5 \\
$18.0 < R \leq 18.5$ & 4 & 1 & 4 & 2 & {\bf ~~0 $\pm$ 3} & 18 & 172 & 3 \\
$18.5 < R \leq 19.5$ & 4 & 1 & 4 & 1 & {\bf ~~0 $\pm$ 2} & -21 & 141 & 2 \\
$19.0 < R \leq 19.5$ & 10 & 3 & 10 & 2 & {\bf ~~0 $\pm$ 5} & 5 & 323 & 5 \\
$19.5 < R \leq 20.0$ & 15 & 3 & 22 & 5 & {\bf ~-7 $\pm$ 8} & -452 & 555 & 1\\
$20.0 < R \leq 20.5$ & 25 & 5 & 34 & 6 & {\bf ~~-9 $\pm$ 11} & -607 &
 731 & 2 \\ 
$ All ~~R$ & ~ & ~ & ~ & ~ & {\bf ~-18 $\pm$ 16} & ~ & ~ & {\bf -2}\\
\hline
\multicolumn{9}{c}{NGC~3656}\\
$R \leq 18.0$ & 8 & 5 & 7 & 4 & {\bf ~~1 $\pm$ 9} & 60 & 564 & 10 \\
$18.0 < R \leq 18.5$ & 3 & 1 & 4 & 1 & {\bf ~-1 $\pm$ 2} & -52 & 127 & 1 \\
$18.5 < R \leq 19.0$ & 7 & 2 & 4 & 1 & {\bf ~~3 $\pm$ 3} & 157 & 197 & 6 \\
$19.0 < R \leq 19.5$ & 8 & 2 & 10 & 2 & {\bf ~-2 $\pm$ 4} & -140 & 261 & 2 \\
$19.5 < R \leq 20.0$ & 21 & 5 & 23 & 5 & {\bf ~~-2 $\pm$ 10} & -107 & 637 & 8 \\
$20.0 < R \leq 20.5$ & 29 & 5 & 35 & 7 & {\bf ~~-6 $\pm$ 12} & -407 &
 755 & 6 \\ 
$ All ~~R$ & ~ & ~ & ~ & ~ & {\bf ~-7 $\pm$ 19} & ~ & ~ & {\bf +12}\\
\hline
\multicolumn{9}{c}{Arp~299}\\
$R \leq 18.0$ & 8 & 5 & 7 & 4 & {\bf ~~1 $\pm$ 9} & 60 & 564 & 10 \\
$18.0 < R \leq 18.5$ & 8 & 3 & 4 & 1 & {\bf ~~4 $\pm$ 4} & 260 & 256 & 8 \\
$18.5 < R \leq 19.0$ & 4 & 1 & 5 & 1 & {\bf ~-1  $\pm$ 2} & -30 & 113  & 1 \\
$19.0 < R \leq 19.5$ & 15 & 4 & 10 & 2 & {\bf ~~5 $\pm$ 6} & 297 & 386 & 11 \\
$19.5 < R \leq 20.0$ & 19 & 4 & 23 & 5 & {\bf ~-4 $\pm$ 9} & -232 & 584 & 5 \\
$20.0 < R \leq 20.5$ & 27 & 2 & 35 & 7 & {\bf ~-8 $\pm$ 9} & -532 &
 567 & 1 \\ 
$ All ~~R$ & ~ & ~ & ~ & ~ & {\bf ~-3 $\pm$ 17} & ~ & ~ & {\bf +14}\\
\hline
\end{tabular}
\caption{ \scriptsize{Statistics on the number of EO in terms of	
counts of expected background galaxies, in intervals of 0.5 $R$ magnitudes. The meaning of
the columns is as follows $\rightarrow$
N: Number of detected EO; $\sigma$: Error in the detected EO counts [($\sigma_{\rm Poisson}^2$ + $\sigma_{\rm field}^2$)$^{0.5}$];
N$_{\rm e}$: Number of expected background galaxies according to Metcalfe et
al. (1991); $\sigma_{\rm e}$: Error in the expected number counts
(Poisson + field-field variations);
dN: Difference between detected and expected number counts;
$\sigma_{\rm d}$: Error in dN ($\sigma$ + $\sigma_{\rm e}$); dN~deg$^{-2}$~mag$^{-1}$:
Difference between detected and expected number counts, per
square degree and 0.5 magnitude;
$\sigma_{\rm d}$~deg$^{-2}$~mag$^{-1}$: Error in dN~deg$^{-2}$~mag$^{-1}$;
N$_{\rm max}$: Maximum number of dwarf candidates (dN + $\sigma_{\rm d}$). For
the last row in each field, dN is the sum of dN$_{\rm i}$ over all the R
intervals, and $\sigma_{\rm d}$ is $(\Sigma \sigma_{\rm d,i}^2)^{0.5}$}}
\end{small}
\end{center}
\end{table*}

Three factors contribute to uncertainties in the galaxy number counts:
Poisson noise, the spatial distribution or clustering properties of
galaxies, and the magnitude error. Following Roche et al. (1993), for
a square field of side-length $\theta$ (degrees), with N galaxies
observed in it and a two point angular correlation function
$\omega(\theta)$ amplitude A, the field-to-field variance in N is
computed using:
\begin{equation}
\sigma^{2}=N + 2.24 N^2 A ~\theta^{-0.8}
\end{equation}  
where a $\theta^{-0.8}$ power-law was assumed for $\omega(\theta)$
(recent results agree very closely with this value; Connolly et
al. 2002), and A has been obtained from Figs.~4 (B band) and 5 (R
band) of Roche et al. (1993) (a conversion log(A(R))=log(A(B))$-$0.5
was applied). These errors were calculated for both the detected (EOs)
and the expected (Metcalfe's) counts: $\sigma$ and $\sigma_e$
respectively, in Table~3. Systematic magnitude errors may also cause
changes in the galaxy number counts. Such zero-point errors relate to
galaxy count errors through the slope of the log[N(m)] distribution
(see Huang et al. 1997 for details). In paper~I, the uncertainty in
the zero point is estimated to be $\pm$~0.1~mag, a value that causes
number count errors that are much smaller than the field-to-field
variations. Metcalfe et al. also reports on a negligible contribution
of the systematic magnitude error to the overall uncertainties.

Table~3 shows the probable number (dN) and maximum number (N$_{\rm max}$)
of TDG candidates obtained by following the procedure described
previously. We note that in some fields -- NGC~520 and Arp~141 --
there are a few more objects than expected from background counts, in
bins 1-2 magnitudes brighter than the completeness limit (i.e. in the
range R=18.5-19.0). NGC~772 shows a $2\sigma$ detection in the same
range, but one $2\sigma$ deviation in a sample of 36 points would be
expected just from Poisson statistics and therefore it is not
statistically significant. The fact that no discrepancies are found
when the total excess of observed counts over background counts is
computed (over the entire magnitude range: last row in each field of
Table~3.), indicates that the TDG population must be very small in
comparison to background objects of comparable size and brightness. In
other words, our results rule out that a {\it typical} galaxy merger
forms 10 or more TDG. Individual cases may produce that quantity, but
not more than {\it a few} -- if any-- TDG can be expected from a
typical galaxy merger.

\section{Conclusions}

Our results do not support evidence for a significant population of
dwarf galaxies around strongly interacting galaxies: the total
probable number of TDG is a small negative number in all galaxy fields
except NGC~772. In this latter field, only one magnitude bin shows a
$2\sigma$ detection of TDG, which is statistically
insignificant. Tidal condensations may turn into bound systems
abundantly, but as self-gravitating, kinematically detached entities,
we expect only a few TDG per collision to be formed. The value
indicated by our results is $\sim$~1 TDG per merger, although as many
as $\sim$~10 TDG cannot be ruled out in individual cases. For three of
the fields -- namely NGC~520, NGC~3226/7 and NGC~3656 -- TDG
candidates have previously been reported (see Balcells 1997; Hibbard
\& van Gorkom 1996; Mundell et al. 1995), but no more than two per
interacting pair and always associated with tidal tails. This result
is also consistent with previous H{\sc i} and optical observations of
six interacting systems (van der Hulst 1979; Hibbard \& van Gorkom
1993; Hibbard \& van Gorkom 1996; Hibbard et al. 1994), where five
dwarf galaxy candidates were reported.

Our results agree with simple theoretical models (Elmegreen et
al. 1993), which predict that only the outermost material will gain
enough angular momentum and energy to become an independent dwarf
galaxy. Most of the bound features along tidal tails found in computer
simulations (e.g. Barnes \& Hernquist 1992) and a substantial fraction
of the luminous condensations reported in the literature
(e.g. Weilbacher et al. 2000) are therefore expected to be short-lived
structures: truly self-gravitating dwarfs that are the result of
galaxy mergers would not be frequent.

\begin{acknowledgements}
The Isaac Newton Telescope is operated on the island of La Palma by
the Royal Greenwich Observatory in the Spanish Observatorio del Roque
de los Muchachos of the Instituto de Astrof\'\i sica de Canarias.
This research has made use of the NASA/IPAC Extragalactic Database
(NED) which is operated by the Jet Propulsion Laboratory, California
Institute of Technology, under contract with the National Aeronautic
and Space Administration.  We thank our referee, J.E. Hibbard, whose
valuable comments have certainly contributed to improve and clarify
this paper. This work has been partially supported by the Spanish DGC
(Grant No. AYA2001-3939). EDD is grateful to the European Union {\it
Research Training Network} Program for its support.
\end{acknowledgements}

\end{document}